# Hidden Signals in Language: Inferring Sensitive Attributes from Reddit Comments Using Machine Learning

**Anay Agarwalla**
anay.agarwalla@gmail.com
**Simeon Sayer**
ssayer@fas.harvard.edu

## *Abstract*

Sensitive attributes are legally protected characteristics that should not be used to discriminate. Careful steps have been taken to minimize the risk of human bias regarding these fields, such as race and age. Large language models (LLMs) are similarly trained not to attempt to infer these aspects. However, just because they shouldn't, doesn't mean they don't. Using chat-like text fragments from authors tagged with sensitive attributes (e.g., MBTI personality, country of origin, gender), a model can often classify these attributes better than a naive guess, with results depending on the combination of subject matter and attribute. The text data from these comments is converted into numerical representations using embedding models, which are then used to train relatively simple classifiers such as logistic regression and decision trees. This study's results show that even these lightweight models can detect statistically significant signals associated with sensitive attributes in user-generated text. The results show that demographic traits such as gender and age are more readily predictable, whereas personality traits are expressed more subtly and depend more heavily on context. Predictive performance varies across online Reddit communities, with some subreddits consistently revealing attributes, while others show high variability depending on the trait being analyzed. These findings indicate that language contains latent identity signals that users may not intend to disclose but are nevertheless detectable through computational methods, and imply that more complex language models may have an inherent, greater capacity to infer sensitive attributes. This raises important concerns about privacy, bias, and the potential misuse of inferred personal information in AI systems. We call for increased transparency, stronger safeguards, and careful policy consideration for future LLMs.

## II. Introduction

Bias is difficult to remove from both human populations and AI models because, often, such biases are subliminal. These biases make surveys an unreliable measure of whether an entity is biased. Many of us don't like to think of ourselves as biased, but we may show it through our actions subconsciously. AI assistants seem to act similarly to the humans that trained them. If asked to guess a user's age, for example, they can dodge the question, refuse to answer, or claim that they are incapable of a response. While these responses appear to show that the model will not rely on assumptions or stereotypes in its logic, they say nothing about what the subliminal model may be doing.

While bias is generally undesirable, specific importance attaches to attributes that are legally protected. These are known as sensitive attributes and cover properties against which discrimination is not permitted: race, gender, age, and personality type are private characteristics that should not be inferred.

While large language models, such as ChatGPT, may claim they do not attempt to predict these attributes, the structure of text processing in these systems makes it difficult to guarantee this. When LLMs convert text into tokens and vectors, linguistic patterns can be captured that correlate with demographic or psychological traits. These signals may exist even when a user does not explicitly mention identifying information.

In this study, we show that even a basic machine learning model can accurately infer sensitive attributes from Reddit comments. If such inferences are possible using a small model and limited resources, it implies that larger language models may also be capable of implicitly making similar predictions. This possibility raises concerns about privacy, bias, and the possible misuse of inferred personal attributes in AI systems.

## III. Background

There is a longstanding interest in both the role of sensitive attributes, such as personality type, and in their ability to manifest in unrelated content, such as online posts. For example, Gjurkovic et al. [1] created a list of Reddit users, and by scraping their self-identified personality types as expressed on one side of the website, such as the MBTI subreddit, they were able to compile a large dataset that explored each user's written content on all other areas of the website. For illustration, a user who identifies as ENTJ in one section had their entire Reddit comment history compiled, so that individual's views in the politics subreddit are documented and associated with their personality type.

This dataset has been used to study the relationship between text-based content and some sensitive attributes. For instance, Jukić et al. [2] used the PANDORA dataset to predict the topic discussed by users in isolation, given context about the person discussing it. This allowed correlations to be drawn; for example, those known to be assertive were far more likely to be discussing politics. This work relates to this paper and appears to perform the inverse task: using personality to infer the topic of discussion, rather than using the topic of discussion to predict personality.

Such data was also used to train models to exhibit a consistent personality by using the entire Reddit history of an author with a known personality type as evidence of that type. This, too, is useful evidence that existing papers have found that a Reddit user's history exhibits a consistent personality fingerprint, meaning it is still viable to infer personality traits from any given comment made by a user. Part of this model involved a Transformer optimally inferring its own personality, which was acknowledged to be weak. That is, when a model responded to a user, it could not reliably tell its own personality in a way that allowed it to continue in the same style. This points to the difficulty of this task and to possible areas for future application of this paper.

Finally, Li et al. [3] found that using a custom dataset that included references to PANDORA, LLMs were comparable to human psychologists in classifying personality type. This also increases the credibility of MBTI predictions regarding NLP and the reliability of language models in understanding the nuances of personality from text. However, human annotators are not infallible, and Li removes what is considered 'noisy data' from the analysis pool. This study makes no such distinction and asserts that all text could provide readily vectorizable fingerprints of sensitive attributes, regardless of discussion topic.

## IV. Dataset

As outlined in the background section, PANDORA and the use of MBTI self-identification in diverse Reddit posts provide a uniquely useful context for analyzing bias and identifying sensitive attributes. It is, therefore, a good candidate for testing whether the AI can pick up on context clues even when these attributes are not directly given.

The PANDORA dataset is structured to minimize duplicated information. As such, a separate file links the full information for each author, held in a CSV file called 'author_profiles', and the comments, held in 'comments.' Reddit usernames are unique, so they serve as a foreign key across all the files. For each author that had the appropriate information identified and stored in

author_profiles, namely MBTI and gender, every comment from a specific subreddit that fulfilled the criteria had its text submission extracted. The chosen subreddit varied based on the diversity of its users, and the overall number of comments exceeded a minimum threshold.

By gluing these files together in Pandas [4], a Python library, specialized subdata were extracted for testing. The sensitive attributes against which these were balanced varied. For instance, in some cases, it was possible to have it that the gender of the comments was an even 50:50 split between those who identified as Male and those who identified as Female. While this ensured a fair prediction regarding gender, it did not guarantee that a diverse representation of personality groups was present in this selection. For instance, a subreddit about partying may have a disproportionately large ratio of extroverted individuals, despite being evaluated for a 50:50 gender split. Forcing the dataset to be balanced with respect to introversion may not be possible due to small sample sizes; therefore, the introversion aspect may be artificially easy to predict for this sub-dataset.

Finally, PANDORA was able to identify authors based on a variety of traits that were all self-identified as part of their user flair. This extends to a mixture of other attributes that may be considered sensitive, such as country of origin. While interesting, the number of users who submitted all possible flairs was so small that it diminished the usefulness of doing so. Therefore, MBTI, gender, country, and age were selected as the only required attributes. Still, in some cases, sufficient authors in a specific subreddit had also volunteered additional information, making further analysis possible.

## V. Methodology and Models

Vectorization of text-based language can be roughly defined as converting language into numbers. This has been an area of extensive research, and there are many relevant pre-existing tools to handle such processing. To enable the application of various classification techniques, it was necessary to determine the optimal methodology for translating Reddit comments from the PANDORA dataset, including known sensitive attributes. By comparing the performance of the combinations of vectorizer and classifier, an optimal combination was found and applied to the data for analysis.

Many different vectorizers for the text were tested. Ultimately, All MiniLM L6 V2 [5] was used due to its well-documented capture of sensitive attributes [6] and its flexibility regarding output vector size. This allowed for a fine-tuning-style approach to adjusting the model vector sensitivity. Additionally, it was broadly trained on online discussion content, such as Reddit and Stack Exchange, meaning it has instinctive familiarity with PANDORA-style inputs.

Once a reliable vectorization strategy was selected, the question arose as to which attributes the model should predict against, and what kind of content was most valuable for making predictions.

The amount of data available in PANDORA was too large to exhaustively search all combinations of attribute and subreddit. Therefore, rather than searching every possible combination, an iterative approach was developed. Initial experiments were conducted on progressively larger subsets of the dataset to identify which subreddit and trait pairings yielded meaningful results. This process allowed computational resources to be directed towards promising combinations, while avoiding unnecessary pairings that produced near-baseline results. Future experimentation may find benefits in analyzing all possible combinations, but due to the sheer size of the dataset, this was impractical for this study. The iterative search process for generating topic/trait recommendations for investigation is deterministic, which complicates the analysis of the 'worst' pairings. All of the evaluated pairings were selected as being most viable, so comparison between subreddits and traits effectively compares the 'worst of the best' (see VI. Results).

After identifying optimal working configurations through these runs, the final evaluation stage scaled the experiment to over one million vectorized Reddit comments across the subreddits and attribute targets selected by the iterative search. This approach provided a balance between the computational power used and the dataset's coverage while still allowing patterns between linguistic signals and sensitive attributes to emerge.

Care was taken during the iterative process to avoid running evaluations on extremely unbalanced attribute populations. This would lead to unbalanced results, where, for instance, if a subreddit's population was largely extroverted, this may result in a naive approach to predicting extroversion, which is always true, despite the dataset being balanced in regard to gender. While the extremes of this were avoided, the necessity to balance against all possible values was prohibitive and forced the shrinking of the dataset.

When evaluating the raw accuracy of the model, questions arise about what that value represents and its real relationship to model performance. By ensuring equal representation, the model was prevented from relying on obvious imbalances, thereby reducing the number of usable samples. This may also have removed some natural correlations within the data; for example, a discussion in the subreddit *r/CollegeBasketball* may be overwhelmingly populated by those who identify as 'under 25' on their scraped flair. By forcing the reduction of the data to allow for a comparison of under-25s and over-25s writing style within this subreddit, the maximum accuracy achievable is significantly reduced compared to that of a naive model that may be able to guess the common attribute. This distinction emphasises that the primary goal of this research is not to achieve perfect classification, but to determine whether or not sensitive attributes can be inferred directly from text.

This was also reflected in the model selection. An advanced custom transformer model would likely achieve much higher performance. Still, the purpose of the experiment is to demonstrate that biases can be relatively easily detected by simpler models within the neural networks that enable chatbots. For this reason, a decision tree and a logistic regression model were used, as the operations that enable their training can be easily mimicked by backpropagation during the chatbot's training. This reinforces any conclusions drawn from the experiment, as it is more plausible that similar behavior is occurring subconsciously in an LLM.

Further to this point, even a relatively weak performance in a forced binary environment such as *r/CollegeBasketball* is an indication that the model is identifying consistent cues from text vectorization, which in turn point towards sensitive attributes. There is an additional margin of error that prevents 100% accuracy, due to the inherently self-submitted nature of the Reddit tags scraped from each account. Online identities are not always completely reliable because users may misrepresent aspects of themselves, exhibit different behavior across contexts, or have multiple users share the same account. Because of this, an extremely high accuracy is both unlikely and could suggest overfitting rather than genuine pattern recognition. For this reason, F1 scores are used in conjunction with accuracy, but even these are not perfect, as they are effectively restricted to evaluating responses of binary systems and to diverse category clusterings with uneven population distribution, such as personality, which are not readily evaluable. Therefore, the most appropriate compromise to observe the effectiveness of a model's output is to compare the difference between the model's solution and that of the naive method of always guessing what the most common group is. This is described as a 'lifted' F1 score and ensures that unwarranted insight is not attributed to models simply for siding with what the majority of the data already represents.

Prior research (see background) strongly indicates that it is not always intuitive how personality, age, or gender could be embedded in a short text snippet, but background research and the experiment in this paper demonstrate that they can (see below). Additionally, for a given subreddit, the attributes generated from these comments are not always trivial, and therefore, an exhaustive search was not possible within the limits of responsible LLM use. However, the recommendation function discussed earlier in this methodology was used to prove that the most exposed components of a given submission are not always directly associated with the subreddit's material, as the results will demonstrate.

# VI. Results and Discussion

The methodology was implemented in a Python environment and yielded a variety of results based on the search-based approach to trait content matching. The highlights of the model findings are presented here, but further interrogation of specific trends may yield additional insights. An important observation from the evaluation process was that the relationship between the discussion topic and the predictability of the sensitive attribute was not always intuitive. In several cases, when the model was run across subreddits that appeared to be closely tied to particular traits, such as those focused on feminism, it surprisingly didn't yield strong predictive performance for related attributes, such as gender. Conversely, some subreddits produced unexpectedly strong accuracy rates for specific attributes that felt very detached from the discussion material. This suggests that trait-related cues can appear indirectly, such as through tone or reasoning patterns, rather than through obvious topic associations. It also validates the effectiveness of the methodology, as these surprising results were readily uncovered, despite not being predictable.

| Trait | macro_f1_lift |
|---|---|
| is_female | 0.254325 |
| country_us | 0.247760 |
| age_under_25 | 0.241801 |
| perceiving | 0.226335 |
| thinking | 0.225098 |
| introverted | 0.224533 |
| intuitive | 0.204034 |

**Fig. 1:** Overall Trait Performance as measured by F1 lift

Using macro_F1_lift to accurately capture model improvement over the naive baseline (see methodology), the average performance of the traits across all discussion topics was calculated. As shown in Figure 1, gender (is_female) achieved the highest average macro_f1_lift, indicating that it was the easiest to predict across all subreddit types. Country_us and Age_under_25 also performed strongly, suggesting that these attributes leave linguistic signals in the text that the model could detect. Interestingly, the four MBTI traits achieved the lowest average predictive performance over the baseline. This suggests that personality-related traits are harder to predict reliably from Reddit comments than demographic traits, possibly because they leave more subtle linguistic cues. At the same time, the MBTI dimensions are fairly close in performance, suggesting that the personality traits are expressed in the text but in weaker ways that depend more on context.

The difficulty of predicting certain attributes also varied across different subreddits. For instance, introversion was more difficult to predict within *r/teenagers*, where users typically communicated using slang, humor, and short responses. In these contexts, the differences between personality types were less visible, making it harder for the model to reliably identify patterns. Conversely, introversion was significantly easier to predict in more politically focused subreddits, such as a community about Donald Trump. In these spaces, comments were usually longer, more argumentative, and focused on personal viewpoints, leading to clearer differences between users. These results suggest that environments that allow greater self-expression may produce linguistic signals associated with personality traits, enabling more reliable predictions.

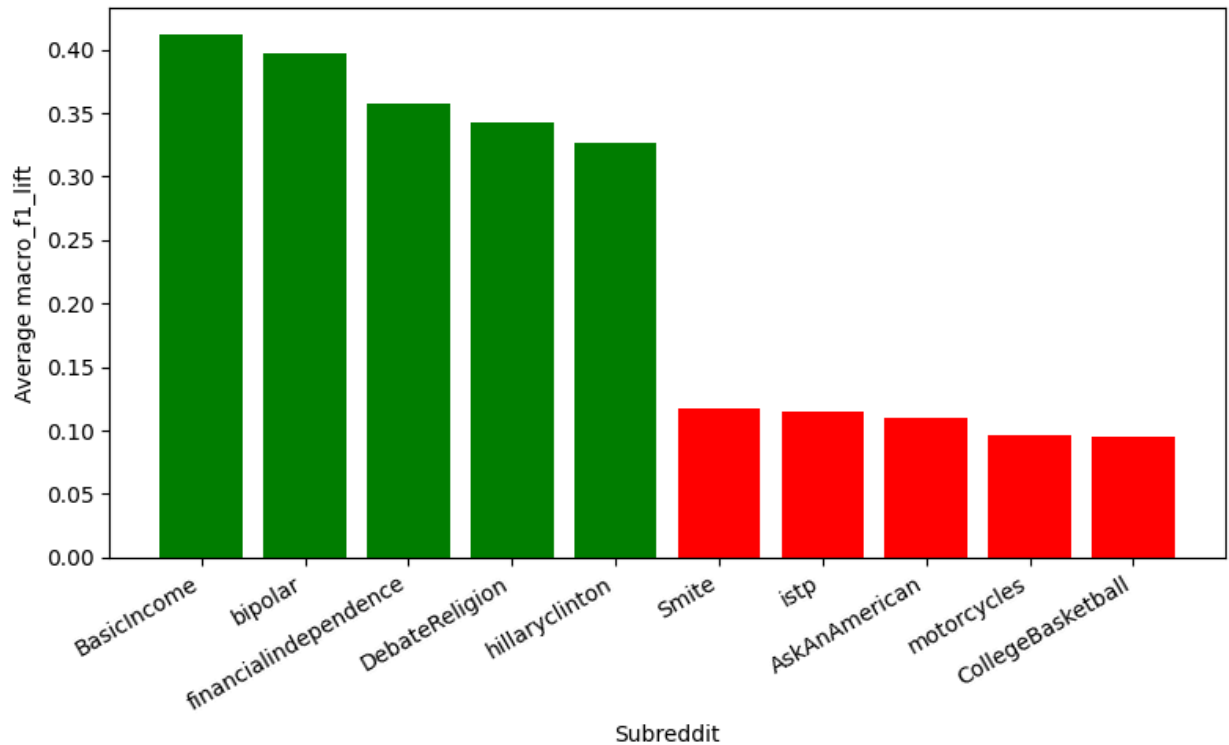

**Fig. 2**: Top (green) vs Bottom (red) Subreddits by Predictive Strength - Average of all macro_F1_lift for all predictions for these subreddits.

Figure 2 shows the highest and lowest-performing subreddits across all evaluated trait combinations. These rankings were produced after the iterative search process identified the most promising subreddit-trait combinations, meaning that even the weakest subreddits shown had already survived an earlier filtering. In other words, the red bars do not represent universally poor-performing subreddits, but rather the weakest results among the most promising candidates. This is important for interpretation, as it means that a subreddit such as *r/hillaryclinton* still shows a macro_F1_lift above 0.1, which indicates that the model found trait-related patterns that enabled it to perform beyond the naive baseline, even in cases that were not among the strongest overall.

This result suggests that explicit discussions about personality do not necessarily make personality traits easier to predict from language. For example, the subreddit *r/ISTP* performs relatively poorly, even though it is centered on MBTI personality types and directly discusses one of the traits being predicted. This could be because users in this community often speak about personality in general terms rather than showing distinctive personal patterns in their own language. In contrast, subreddits about finance, such as *r/BasicIncome* and *r/FinancialDiscussion,* produce a stronger predictive performance. Posts in these communities often discuss personal circumstances, life planning, and the economy, which can reveal demographic information. For example, users may explicitly mention details such as "I'm a single male" or describe their careers and goals. As a result, even though these subreddits are not explicitly about identity or personality, they provided more obvious signals about the people writing the posts.

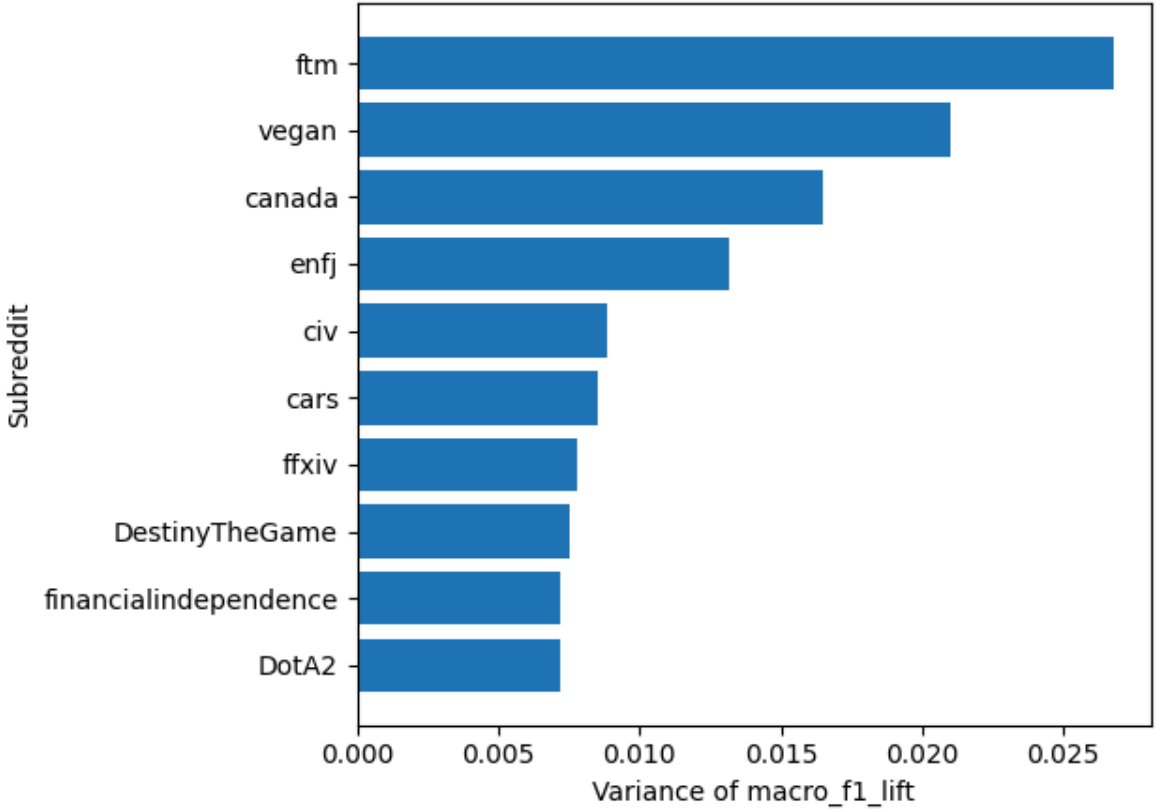

**Fig. 3:** Top 10 Subreddits with Highest Trait Prediction Variance

High variance in predictive performance for a given subreddit indicates that the effectiveness of trait prediction in that subreddit is highly dependent on the trait. Certain traits may produce very strong predictive signals in a subreddit, while in that same community, different traits may be extremely hard to predict. This suggests that sensitive trait detectability is not binary. It is possible, as evidenced by Fig 3, to have text content that consistently exposes one aspect of the user, while not exposing others.

The traits that are exposed are not consistent either. Take the subreddits *r/ftm* and *r/vegan,* which demonstrated a particularly high variance in trait prediction performance. These communities are spaces where discussions are primarily about shared experiences, personal narratives, and identity-related topics. Hence, language used in these communities may contain strong signals related to identity or demographic characteristics but weaker signals associated with personality traits. This explains why some attributes can be predicted relatively well within these communities while others are closer to the baseline.

Figure 4 (below) compares the resulting variance between these two subreddits.

| Trait | r/FTM (macro_F1_lift) | r/Vegan (macro_F1_lift) |
| --- | --- | --- |
| perceiving | 0.337657 | 0.382776 |
| age_under_25 | 0.262881 | 0.154686 |
| thinking | 0.024375 | 0.113626 |

**Fig. 4:** Comparison of the relative performance of high-variance subreddits with shared traits

The fact that the two highest variance subreddits were tested on the same traits may have some relationship, or may simply be happenstance. For instance, two subreddits with the next-highest variance, *r/Canada* and *r/ENFJ,* showed large differences due to weak gender predictability.

As a point of comparison, the highest overall average for predictability, *r/BasicIncome,* has an extremely low variance of 0.00004. It therefore appears that some subreddits are made up of identity communities, such as *r/ftm* and *r/vegan*, where discussions focus on shared identity experiences. These communities tend to produce stronger signals related to demographic characteristics but weaker signals related to personality traits, leading to higher variance. Others are generically revealing as to all aspects of the user, and seem to be ideological or debate communities, such as *r/DebateReligion* and *r/BasicIncome*, where users frequently hold extended arguments or discussions about beliefs. In these environments, the whole personality is reflected in the linguistic style, leading to the model achieving a consistently high predictive accuracy.

On average, a given selected attribute achieved a macro_F1_lift of 0.232. A specific edge case of note is the highest combination, achieved by identifying the 'age_under_25' attribute in *r/bipolar*, which had a macro_F1_lift of 0.443, indicating it is significantly stronger than simply guessing the majority. The weakest combination was the 'thinking' attribute in the *r/ftm* subreddit, which was only marginally better than the F1 score of guessing with the majority, at 0.02. Even in the worst case, however, it is clear that using the vectorization is more useful than a naive guess, and in most cases, significantly better. This offers implications for sensitive attribute detection, especially within complex models.

## VII. Conclusion

The results provide concrete evidence that greater care should be taken in how AI models interact with sensitive attributes. This initial experiment demonstrates that there are many directions for future research.

First, the data used in this study could be expanded beyond the PANDORA dataset to obtain diverse samples of online communication. Since PANDORA relies on self-identified attributes and focuses on Reddit comments, new data sources could include other platforms or text sources to determine whether the same linguistic signals appear across different settings. Second, other sensitive attributes beyond the limited set examined in this paper could be investigated, potentially revealing additional signals embedded in written language. The dataset would also benefit from objective, independently reviewed classification of sensitive attributes, rather than the self-identified approach used by PANDORA. Finally, more advanced machine learning models could be applied to the same problem to determine how much predictive performance improves as computational power increases. This could represent a more advanced vectorizer or a higher parameter classification technique. However, the ultimate goal of this experiment is to demonstrate how easily these trends can be reliably identified, even with noisy data and even on simple, logistic regression-driven models.

The study reinforces the fact that language contains subtle indicators of identity that individuals may not intend to reveal. Even when users do not explicitly mention personal information, their word choice, tone, and reasoning patterns may inadvertently reveal aspects of their background or personality. This proves that machine learning systems analyzing large volumes of text may be able to infer more about users than users expect. These inferences may be used in ways that are at best surprising and at worst discriminatory.

Building on this idea, if relatively simple models can infer sensitive traits from a smaller pool of text data, then modern large language models trained on massive datasets may be even better at detecting these signals. These models may implicitly learn correlations between language patterns and personal attributes during their training, even if they are not designed to predict them. As a result, LLMs may possess knowledge about the user that is not visible to the user, and may even deny having it. Understanding how these models process linguistic signals and expanding model interpretability to better examine the effects of sensitive attributes on the model are, therefore, important areas for future research.

It is worth acknowledging, however, that this ability may not be completely unique to artificial intelligence. Humans also tend to form impressions of others based on written communication, often subconsciously inferring traits such as age from tone or vocabulary. In this sense, AI models may be mimicking a kind of social intuition that people already use when interpreting language. The key difference, and what this paper proves, is that machine learning systems can identify these patterns across much larger datasets and with greater consistency. As a result, the model can outperform human intuition, not because it understands the writer more deeply, but because it can detect subtle correlations across millions of examples. This raises important questions about how such capabilities might be applied, and whether the rules for an artificial intelligence need to be significantly different to that of a human's. While these insights could enhance applications such as personalization or marketing, they also raise concerns about large-scale profiling and both the ethical and legal implications of inferring personal traits from seemingly anonymous text.

Ultimately, as language-based AI systems continue to advance, understanding how these hidden linguistic signals are processed will become increasingly important. The ability of models to infer personal characteristics from everyday communication suggests that language itself contains much more information about identity than is immediately apparent. This capability could eventually open up new areas of research in psychology and human-computer interaction, helping researchers better understand how people express identity through language. Simultaneously, it emphasizes the need for careful consideration of privacy and transparency in AI systems. As machine learning models become more powerful and integrated into online platforms, researchers and policymakers will face the challenge of balancing the benefits of these technologies with responsible safeguards so that existing rights of users are protected.

## VIII. Acknowledgements

I would like to acknowledge the support of Simeon Sayer at Harvard University for his mentorship during this project.